# Ventilation Perfusion Functional Difference Images in Lung SPECT: a linear and symmetrical scale as an alternative to the Ventilation Perfusion Ratio

Robin de Nijs[1], MSc PDEng PhD
Ronan M. G. Berg[1,2,3], MD DMSc
Sofie Lindskov Hansen[1], MSc PhD
Jann Mortensen[1,4,5], MD DMSc

1. Department of Clinical Physiology and Nuclear Medicine, Copenhagen University Hospital - Rigshospitalet, Copenhagen, Denmark
2. Department of Biomedical Sciences, Faculty of Health and Medical Sciences, University of Copenhagen, Copenhagen, Denmark
3. Center for Physical Activity Research, Copenhagen University Hospital - Rigshospitalet, Copenhagen, Denmark
4. Department of Clinical Medicine, Faculty of Health and Medical Sciences, University of Copenhagen, Copenhagen, Denmark
5. Department of Medicine, The National Hospital, Torshavn, Faroe Islands

November 4, 2026

Corresponding author: Robin de Nijs, orcid.org/0000-0002-3908-3013

Email: robin.de.nijs@regionh.dk

Phone: +45-35454011

## Abstract

*Purpose:* Ventilation Perfusion SPECT is important in the diagnostics of e.g. pulmonary embolism and chronic obstructive pulmonary disease. Classical and reverse mismatched defects can be identified by utilizing the ventilation-perfusion ratio. Unfortunately, this ratio is only linear in the ventilation, the scale is not symmetrical regarding classical and reversed mismatches and small perfusion values give rise to artifacts. The ventilation-perfusion (VQ) difference is developed as an alternative.

*Methods*: For both VQ-ratio and VQ-difference a scaling factor for the perfusion is computed, so that voxels with matched ventilation and perfusion (on average) yield zero signal. The relative VQ-difference is calculated by scaling with the summed VQ-signal in each voxel. The scaled VQ-difference is calculated by scaling with the global maximum of this sum.

*Results*: The relative and scaled differences have a scale from -1 (perfusion only) to +1 (ventilation only). Image quality of relative VQ-difference and VQ-ratio images is hampered by artifacts from areas with both low perfusion and low ventilation. Ratio and differences have been investigated in ten patients and are shown for three patients (one without defects). Clinical thresholds for the difference images are derived resulting in color maps of relevant (reversed) mismatches with a (reciprocal) ratio larger than two.

*Conclusions*: The relative ventilation-perfusion difference is a methodological improvement on the ventilation-perfusion ratio, because it has a symmetrical scale and is bound on a closed domain. A better diagnostic value is expected by utilizing the scaled difference, which represents functional difference instead of relative difference.

246 <250 words

## Introduction

The current EANM guidelines emphasize the added diagnostic value of the ventilation-perfusion ratio (V/Q ratio, also coined the V/P quotient) on VQ scintigraphy, which is obtained by measuring the quantities V and Q in every voxel [1]. Ventilation (V) refers to the flow of air into and out of the airways and alveoli, while perfusion (Q) refers to the blood flow through the pulmonary circulation. On this basis, matched ventilation-perfusion defects (V/Q ratio≈1) may easily be distinguished from 'classical' mismatched perfusion defects [2], that is, areas with reduced perfusion and maintained ventilation (high V/Q ratio), of which the former indicates ventilatory disturbance as can be seen in Covid-19 lung disease (3,4,5) and chronic obstructive pulmonary disease (COPD) [6], whereas the latter is observed in pulmonary vascular disease, most notably pulmonary thromboembolism [7]. Furthermore, in some lung diseases associated with intrapulmonary shunting, regional ventilation may be reduced with maintained perfusion, which is associated with a so-called reversed mismatch, with a low V/Q ratio.

The use of the V/Q ratio as an imaging parameter is problematic for several reasons. Firstly, because the part of the intensity scale that is relevant for the V/Q ratio in the case of a reversed mismatch goes from zero to one, while it goes from one to infinity in the case of classical mismatch. This makes the simultaneous visualization of classical and reversed mismatch difficult in practice. Secondly, the voxelwise ratio between V and Q varies linearly with V as the numerator, but - since Q is in the denominator - not with Q, so that artifacts often arise when Q-values become very small. Thirdly, use of the V/Q ratio is not suitable for detecting ventilatory disturbances, as matched ventilation-perfusion defects have the same value as areas with normal ventilation and perfusion. For the same reason, the V/Q ratio does not provide any useful information regarding regional lung function in such instances.

Due to the abovementioned limitations, we posit that the regional V-Q *difference* is a more valid approach for scintigraphic imaging and quantification of VQ relationships. In the present paper, we thus provide the mathematical rationale for a V-Q difference based approach and compare it to the regional V/Q ratio and its reciprocal value.

## Materials and methods

*Patients and scanning protocol*

Patient data from the SECURe Study[5] was utilized. The SECURe Study is a prospective cohort study of patients that have suffered from COVID-19 of various severities. In this study VQ SPECT/CT was performed with $^{99m}$Tc labelled macroaggregated albumin (MAA) as the perfusion tracer and $^{81m}$Kr as the ventilation tracer. The typical administered activities were 150-300 MBq for $^{81m}$Kr (effective dose: 0.1-0.2 mSv) and 40-150 MBq for $^{99m}$Tc (effective dose: 0.4-1.7 mSv). The SPECT/CT protocol started with a 110 kVp, 20 mAs low-dose CT with dose modulation reconstructed in 3 mm axial slices (~1.5 mSv), a 512x512 matrix and 1.37 mm isotropic pixel size, after which the dual-isotope VQ SPECT was acquired (128x128 matrix, 128 angles, 10 s per angle in step and shoot mode, 4.8 mm isotropic voxels, parallel hole Medium Energy Low Penetration (MELP) collimators). $^{81m}$Krypton was inhaled during normal tidal breathing. All scans were performed on a dual-head 16-slice CT Siemens Intevo Bold (Siemens, Erlangen, Germany) with 9.5-mm-thick-NaI(Tl)-scintillation crystals, and the total acquisition time was approximately 11 minutes.

Scatter correction was performed with in total five energy windows. The percentagewise window width was defined relatively to the main photopeak. Since the 140 keV (15% width) $^{99m}$Tc includes down-scatter from $^{81m}$Kr photons, the Triple Energy Window (TEW)-method [8] with broad energy windows [9] for $^{99m}$Tc was applied with widths 15% for the lower scatter (weight 0.5) and 10% for the upper scatter (weight 0.75). The $^{81m}$Kr main window was defined at 191 keV (15% width) and with the lower scatter window set just below with a 6% window width (weight 1.25). The energy windows between the $^{99m}$Tc and $^{81m}$Kr peak were set at 10% and 6%, respectively, because the scanner does not allow overlapping energy windows. The applied scatter correction is similar to the TEW-method applied on the two main peaks of $^{177}$Lu [10]. The acquired data was reconstructed with the iterative Flash3D (3D-OSEM with 4 iterations, 16 subsets, 12 mm Full Width Half Maximum (FWHM) Gauss filter) including collimator detector response modelling and CT-based attenuation correction.

*V/Q ratio and functional difference*

The voxelwise ventilation-perfusion ratio $R$ is defined by

$$R = \frac{V}{kQ}, \tag{1}$$

where $V$ and $Q$ refer to the ventilation and perfusion signal in each voxel, respectively. The scaling factor $k$ is included in order to obtain a ratio of one, when ventilation and perfusion are matched. For low perfusion values the ratio value can quickly become very large. For displaying ventilation defects the reciprocal ratio can be used, but it becomes very large values for low ventilation values. A practical solution of this would be to take the logarithm of the ratio $R$, but it only reduces and does not solve the problem with large values. However, it makes simultaneous display of the ratio below and above one possible, since the logarithm causes the ventilation defect to be displayed as a negative number with otherwise the same scaling as perfusion defects.

The voxelwise functional V-Q difference is defined by

$$D = V - kQ, \tag{2}$$

where $k$ in this case is a scaling factor resulting in a difference $D$ of zero, when ventilation and perfusion are matched. $D$ is both linear in $V$ and $Q$ and does not contain artifacts when $Q$ becomes small. However, a quantitative measure is preferred, so that a threshold for clinically relevant defects can be defined. This measure should not depend on e.g. the camera sensitivity or the amounts of administered tracers. We propose the following measure:

$$D_{\mathrm{rel}} = \frac{V-kQ}{V+kQ} = \frac{R-1}{R+1}, \tag{3}$$

which will be referred to as the relative difference. The denominator $V + kQ$ will be referred to as the total signal. By including the scaling factor $k$, the total signal does not depend on the ratio of the administered tracer amounts, and the relative difference is independent of the administered amount for each individual tracer. The $D_{\mathrm{rel}}$ -measure potentially has a similar problem (division by a very small number) as $R$ in equation (1), if both $V$ and $Q$ are small. For small changes in $V$ or $Q$ from the situation of matched ventilation and perfusion ($V$=$kQ$) the $D_{\mathrm{rel}}$-value equals ($R$-1)/2, which equals half the relative change. For small deviations from $R$=1 the relative difference $D_{\mathrm{rel}}$ and half ln $R$ are identical to second order[1]. An identical relative change in $V$ or $Q$ results in the same absolute value, but it is positive for perfusion defects (mismatch) and negative for ventilation defects (reversed mismatch).

The functional difference $D$ defined by equation (2) has the advantage that it is linear for both perfusion and ventilation. This is not the case for the ratio $R$ and relative difference $D_{\mathrm{rel}}$. Although $D_{\mathrm{rel}}$ has some of the same disadvantages as $R$, it has some important advantages. The scale is limited from -1 for pure perfusion to 1 for pure ventilation. It can express both ventilation and perfusion defects simultaneously, where the value of both types of defects mathematically behave in the same way. In order to keep the scaling properties, the difference $D$ is divided by a scalar and the scaled difference $D_{\mathrm{S}}$ is defined by:

$$D_{\mathrm{S}} = \frac{V-kQ}{|V+kQ|_{\max}}. \tag{4}$$

The maximum of the total signal is chosen, because the mean would include areas in the lungs with low ventilation and perfusion. For displaying purposes, a relevant clinical threshold for the scaled difference can be approximated. Assuming arbitrarily that ventilation-perfusion ratios above two for perfusion defects with unaffected ventilation might be relevant, a clinical threshold of $V/4V_{\max}$ is obtained by assuming that ventilation and perfusion are balanced in the maximum, i.e. $V$=$kQ$. Since the maximum value is used, the optimal threshold value will be lower. By applying 50% of $V_{\max}$ as a representative value for $V$ the threshold becomes $V/4V_{\max} = V/8V = 0.125$. Generally, threshold values $D_{\mathrm{S,th}}$ for perfusion and ventilation defects can be calculated as a function of $R$ with

$$D_{\mathrm{S,th}}(R \geq 1) = \frac{c}{2} \cdot (1 - R^{-1}) \quad \text{and} \quad D_{\mathrm{S,th}}(R < 1) = \frac{c}{2} \cdot (R - 1), \tag{5}$$

respectively. The ratio between representative value and maximum is denoted with $c$. The threshold values for ventilation defects with unaffected perfusion are equal to the threshold values for perfusion defects with opposite sign.

---

[1] $\frac{1}{2}\ln R = r - r^2 + \frac{4}{3}r^3 + \cdots$ and $D_{rel} = r - r^2 + r^3 + \cdots$ with $r = (R-1)/2$

For comparison and displaying purposes we focus on (reversed) mismatches with a (reciprocal) ratio between two and five. In Table 1 the difference values to which these ratios correspond are shown together with the minimum, maximum and the value for a ratio of one for matched ventilation and perfusion. For determining the scaled difference for a ratio of five as an upper limit a $c$-value of 80% of $V_{\max}$ was used as a representative value for $V$.

***Table 1**: Values for ratios and differences for comparison and displaying. For mismatches the upper display limit (R=5) of the scaled difference was determined using a c-value of 80% of $V_{max}$ for V and the lower display limit (R=2) was determined using a c-value of 50%. Limits for reverse mismatches were obtained in a similar way.*

| | ratio $R$ | difference $D_{\mathrm{rel}}$ | difference $D_{\mathrm{S}}$ |
|---|---|---|---|
| minimum | 0 | -1 | -1* |
| reverse mismatch | 0.2 | -0.67 | -0.32 |
| | 0.5 | -0.33 | -0.125 |
| perfect match | 1 | 0 | 0 |
| mismatch | 2 | 0.33 | 0.125 |
| | 5 | 0.67 | 0.32 |
| maximum | ∞ | 1 | 1* |

* These values are calculated without the assumption of balanced ventilation and perfusion, i.e. $V=kQ$, for the maximum of the total signal in equation (4), otherwise these values are equal to -0.5 and 0.5 according to equation (5) with $c$=1.

*Calculation of the scaling factor*

For ratio images the calculation from the scaling factor follows naturally from equation (1). Ignoring defects and assuming that most of the signal is in the lung area, the scaling factor can be computed by

$$k = \frac{\bar{V}}{\bar{Q}} = \frac{\sum V}{\sum Q}. \tag{6}$$

The mean of the voxelwise $V/Q$ is biased [11]. For difference images it is more natural to determine the scaling factor following equation (2) by minimizing the difference in a least squared sense and it is given by

$$k = \underset{\hat{k}}{\arg\min} \sum \left| V - \hat{k}Q \right|^2 . \qquad (7)$$

These approximations are often sufficiently accurate (in the case of few defects), but there are methods to take the defects into account, such as the z-map method by Boussion et al [12] and the iterative subtraction z-map method by Thomsen et al [13]. The method by Boussion et al is based on finding a reference area for which the scaling factor - the $k$-value - is calculated. First, a mask with a threshold of 40% of the maximum intensity is applied on each individual image. Voxels with an intensity within one standard deviation from the mean are kept for each individual image. The reference area consists of the intersection of the remaining voxels in both images.

We applied the method by Thomsen et al. First, the initial $k$-value was calculated by using equation (6) for the ratio images. A mask, which removes all voxels in the total signal image ($=V+kQ$) lower than 10% of the maximum, was calculated using this initial $k$-value. Additionally, the ventilation and perfusion images were also masked with a 10% threshold. The intersection of these three masks was used for masking the ventilation and perfusion images. After applying the latter mask, equation (6) was used for calculating an updated $k$-value. The mask of the total signal image was updated with the updated $k$-value and the process was repeated and stopped at 100 iterations or convergence. The $k$-value for the difference was calculated by an identical procedure, except that equation (7) instead of (6) was used.

The 10% threshold was only used for calculation of the scaling factor avoiding noisy and low signal contributions and focusing on areas with the highest signal. For display purposes, a 3% mask of the total signal images ($=V+kQ$) was used, only when artifacts outside the lungs were present, as was the case for the ratio and relative difference images.

## Results

In order to illustrate the different measures of V, Q and their relation, Figure 1 shows the two quantities and their ratios. The same transversal slice is shown in Figures 1 and 2. In the top right and bottom right the mask for calculating the scaling factor $k$ and the lung mask for the visual removal of artifacts are shown, respectively. An area with a high V/Q ratio is visible in the middle of the top middle image. This is the respiratory tract, where only ventilation is present. In the bottom middle image a high Q/V ratio in the lower part of the image is observed in both lungs, indicating a ventilation defect.

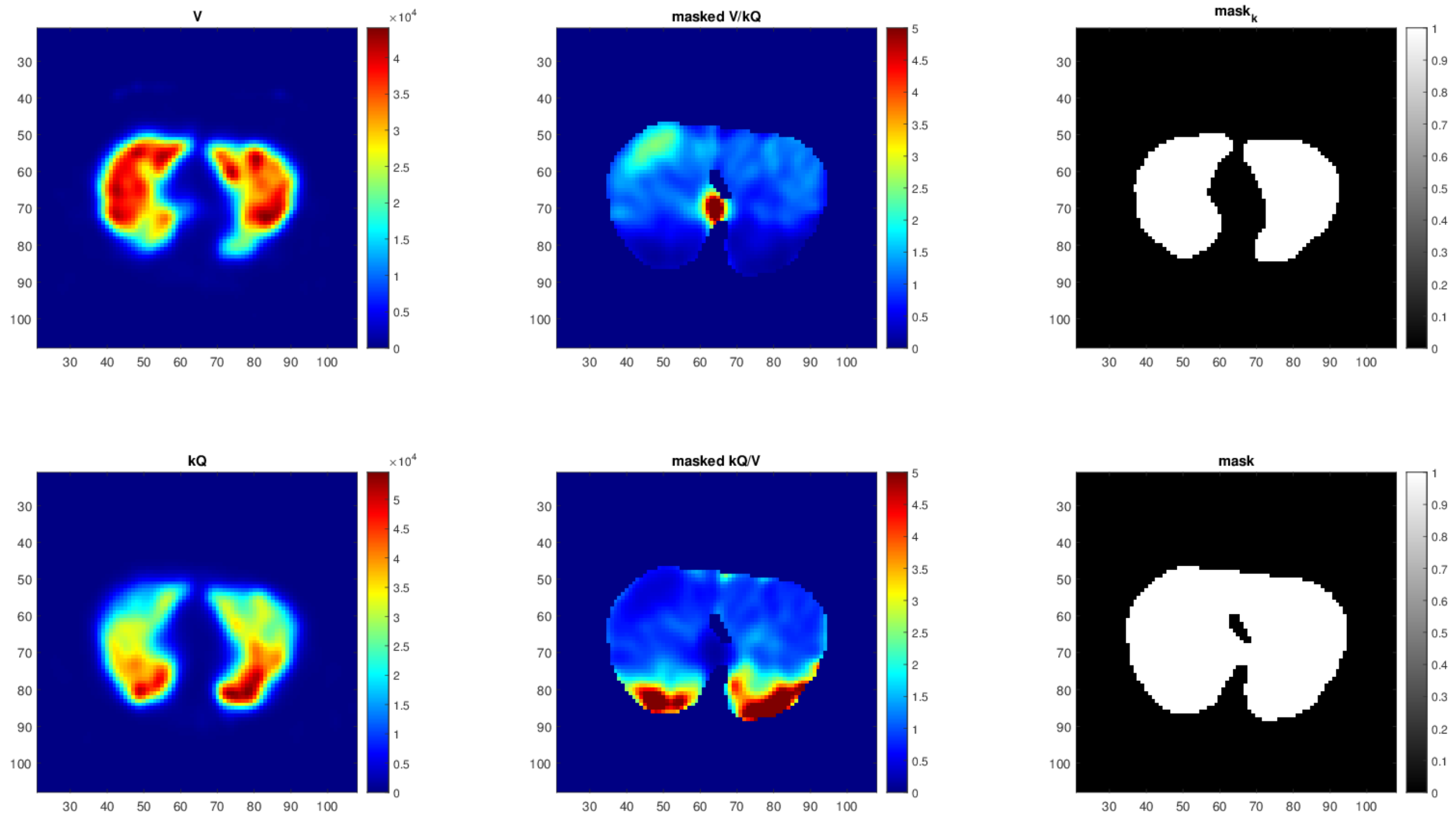

***Figure 1**: The same transversal slice is displayed in all images. Top row: ventilation, masked V/kQ ratio and mask for the calculation of the scaling factor k with a calculated value of 8.37. Bottom row: scaled perfusion, masked reciprocal V/kQ ratio and lung mask for displaying. The maximum of the ratio color scales is five.*

Figure 2 shows (natural) logarithmic ratio images, relative difference images, scaled difference images and the sum image of ventilation and perfusion. The advantage of the scaled difference instead of the ratio or relative difference is illustrated. First, the logarithmic ratio image (top left) shows the necessity of applying a mask, when calculating ratios and the like. Secondly, the masked logarithmic ratio (top middle) and the masked relative difference (top right) look very similar. From the ratio and relative difference images, reversed mismatches areas are visible in both lungs in the bottom of the image, which is also seen on the scaled difference image. The respiratory tract is visible as a "mismatch" in the middle of the masked (logarithmic) ratio and masked relative difference images, but not in the scaled difference images. The scaled difference (bottom middle image) has a different spatial distribution. The bottom right image shows the voxel values with an absolute value larger than 0.125 showing the clinical significance of the previously mentioned areas.

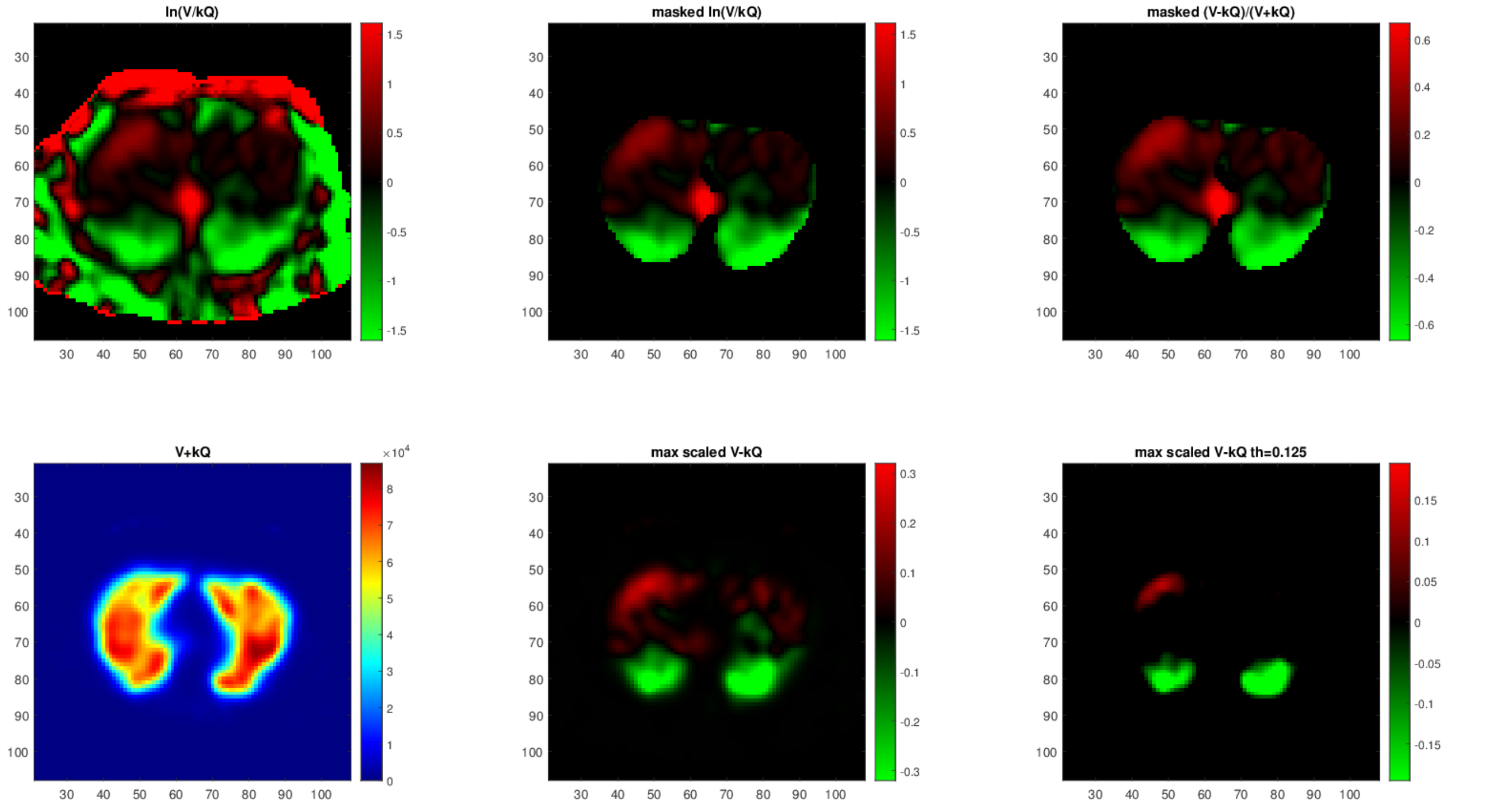

***Figure 2****: The same transversal slice as in Figure 1 is displayed in all images. Top row: the natural logarithm of the ratio, the masked logarithmic ratio, the masked relative difference image. Bottom row: sum image of ventilation and perfusion, scaled difference image with limits ±0.32, scaled difference image with a threshold of ±0.125. The intensity scale in the latter image has limits ±0.195 after subtraction of the threshold value utilizing the entire color scale. The minimum and maximum of the all green-red color scales correspond to a ratio R of 0.2 and 5, respectively, see Table 1. The numerical value of the k-value for the difference images is 8.03.*

In Figures 3 and 4, some relevant coronal slices are shown for the same patient as in figure 1 and 2. In this case, positive and negative ventilation perfusion ratios/differences are shown separately, and the images are fused with a low dose CT. Figure 3 shows mismatches with low perfusion compared to ventilation as is the case in pulmonary embolism. The respiratory tract is clearly visible, as is a mismatch in the upper half of the right lung. The ratio and relative difference images show a problem, where the signal is not contained inside the respiratory tract resulting in a visual overexposure effect. This problem is slightly reduced for the relative difference, but only the scaled difference image does not have this problem. The maximum of the mismatch in the upper half of right lung appears at another position in the scaled V-Q difference.

Figure 4 shows reversed mismatches, and here it can be seen that the ratio (left) and relative difference images (middle) show enlarged areas of reversed mismatch. In this case, the right lung would be diagnosed having severe ventilation defects, while there is actually only a minor mismatch on the inner side of the bottom right lung as shown in the scaled difference image (right). The relative difference image does not really reduce this problem. The left lung has three areas with a severe reversed mismatch as shown on the right image. Ratio and relative difference images show mismatches on the edge of the lungs, while the scaled difference image show the reversed mismatches inside the lungs.

In Figure 5, ventilation images of a patient without ventilation or perfusion defects are shown. There should be no mismatches, apart from the respiratory tract, which is the case for both the relative and the scaled V-Q difference. The V/Q ratio image shows an oversized respiratory tract and several hot areas in the lungs. That problem is significantly reduced by the relative difference, but the scaled functional difference outperforms the latter.

Figure 6 shows images of a patient with a large perfusion defect in the bottom left lung as can be seen in the bottom right image. Intensity scales for the ratio and the relative difference are adjusted to a maximum value of 10 and 1, respectively. The problem of the non-closed domain can be seen by comparing the first two images of the top row. Ratio values can be very large for small perfusion values, while they are just close to one for the relative difference images. However, only the scaled functional difference image shows a well contained area of the perfusion defect. Notably the scaled difference intensity is not so high, because there actually is slightly less ventilation in the area with the perfusion defect, as can be seen in the bottom middle image.

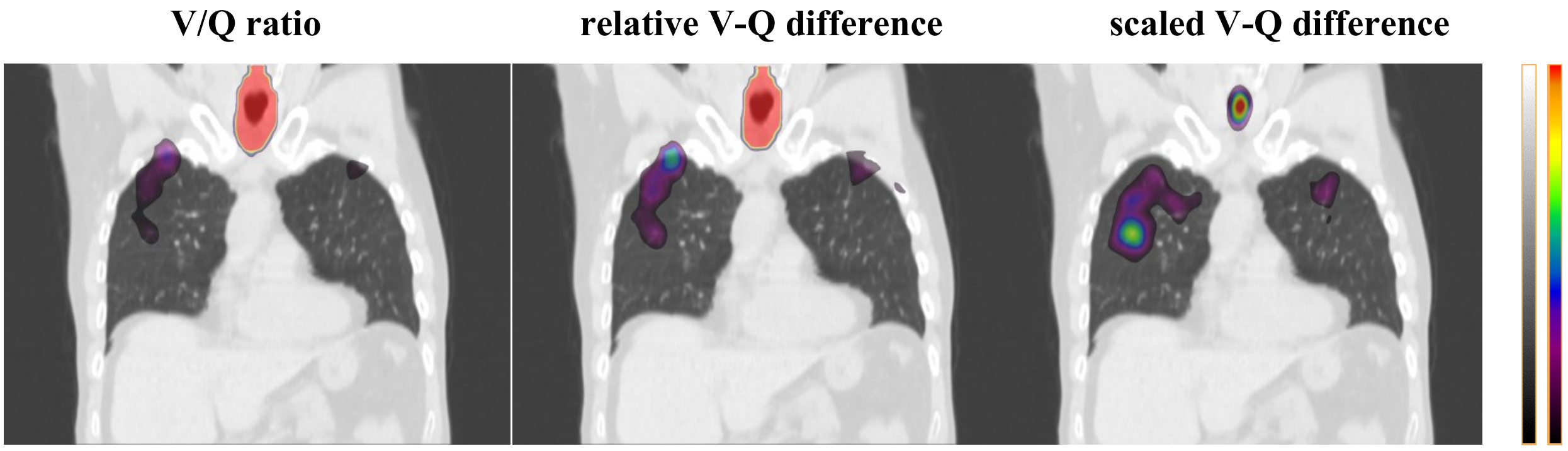

***Figure 3****: Coronal slices fused with a low dose CT. A positive difference scale for mismatches is shown. The CT color scale goes from -1400 to 200 HU as common for lung CT. The rainbow color scale corresponds for the scaled difference ratio images (middle) to a minimum ratio of two and a maximum ratio of five (from 0.33 to 0.67). The scaled difference images (right) have a color scale from 0.125 to 0.32. In this way the boundaries of the SPECT color scales are compatible.*

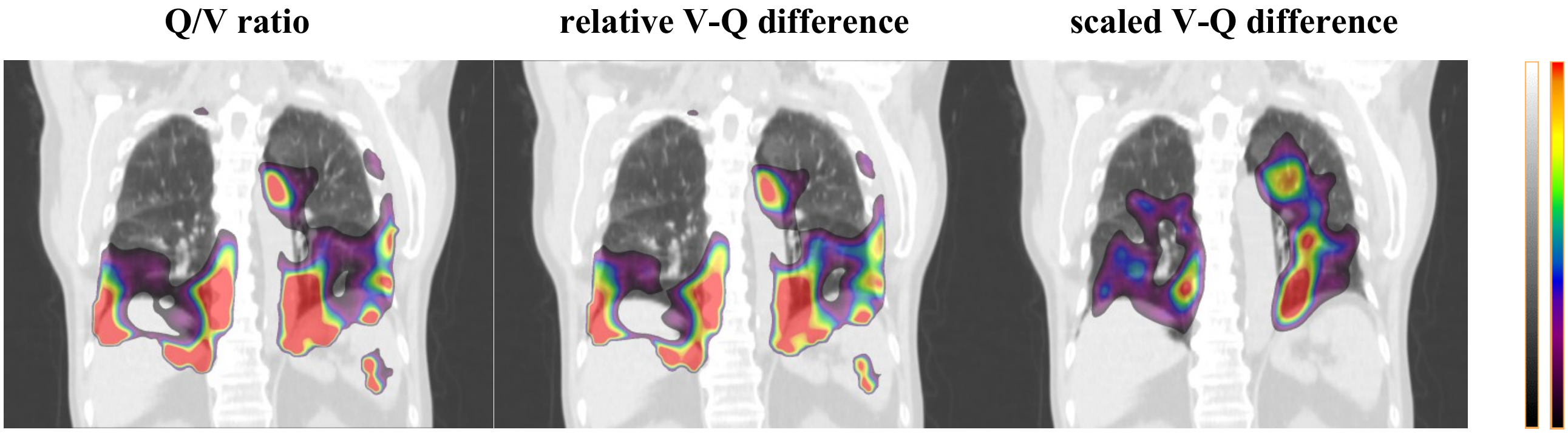

***Figure 4****: Coronal slices with a low dose CT. A positive difference scale for reversed mismatches is shown. The CT color scale and rainbow color scale are the same as in Figure 3.*

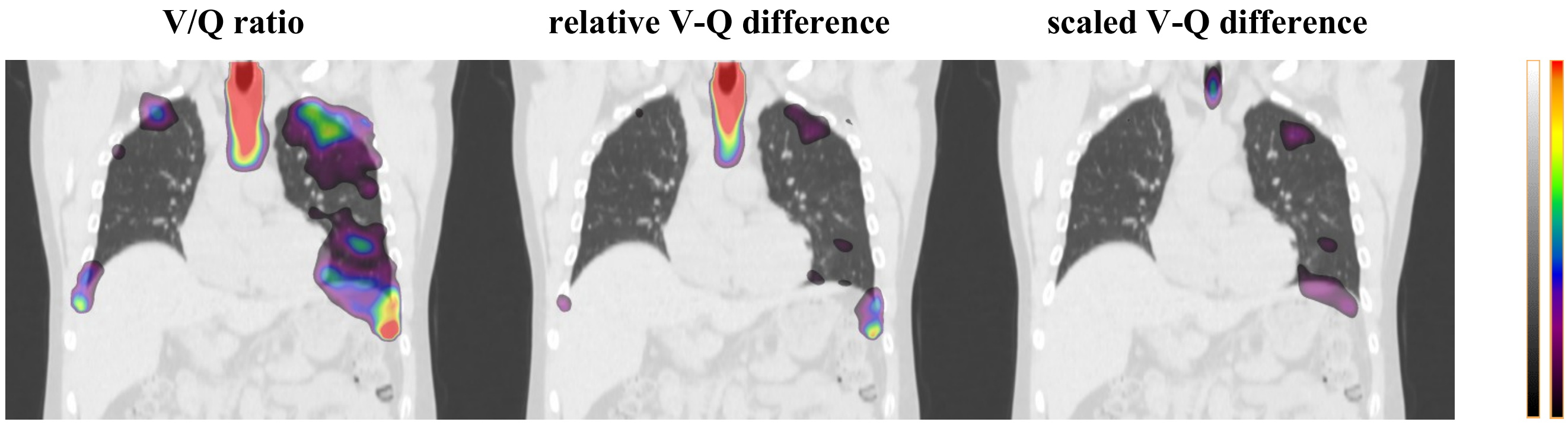

***Figure 5****: Mismatch images for a patient without ventilation or perfusion defects. Coronal slices with a low dose CT. A positive scale for mismatches is shown. The CT color scale and the rainbow color scale are the same as in Figure 3.*

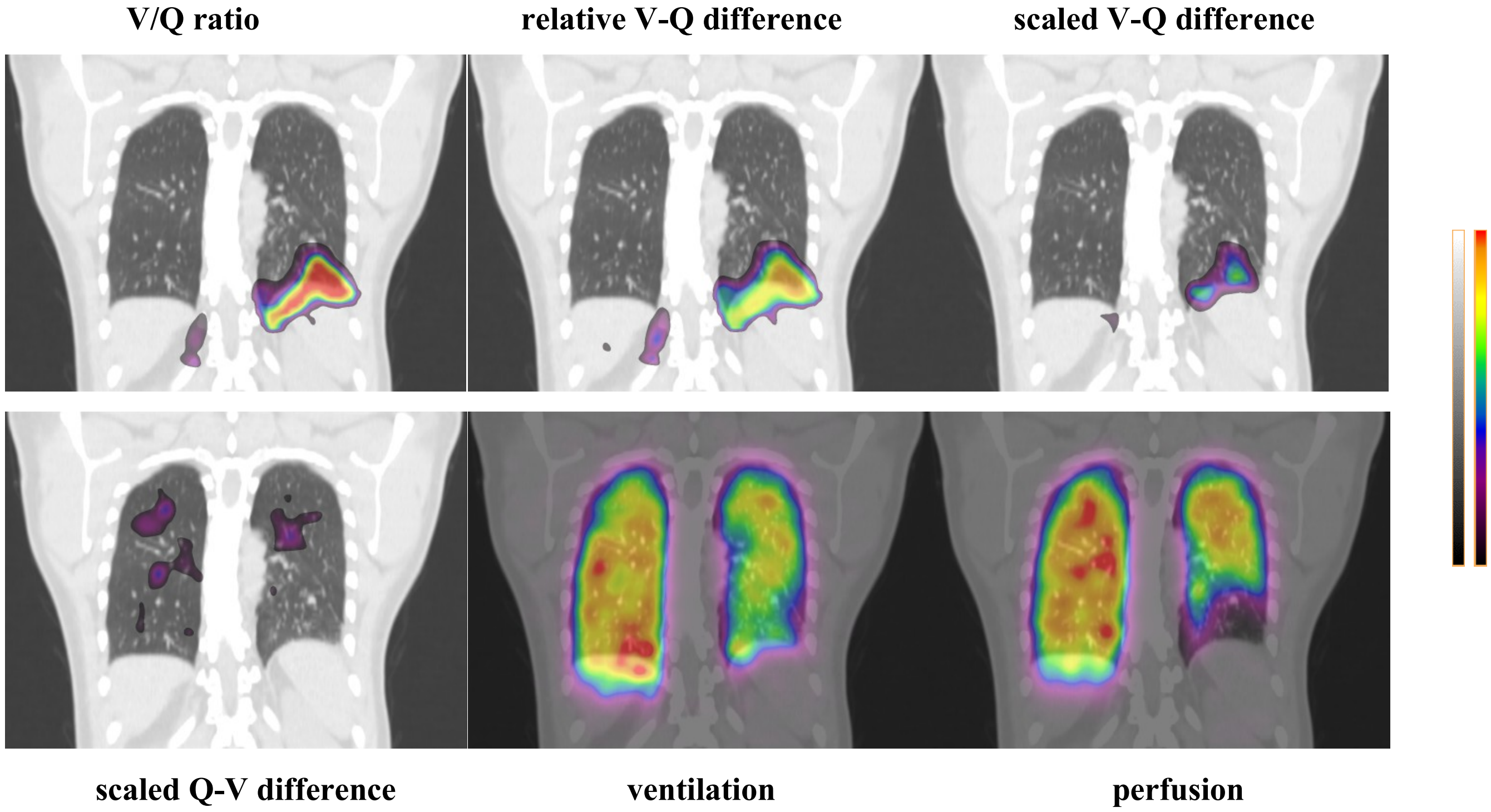

***Figure 6****: Images of a patient with a large perfusion defect in the bottom left lung (pulmonary embolism). Coronal slices with a low dose CT. A positive scale for mismatches is shown. The CT color scale is the same as in Figure 3. Top row: V/Q ratio (scale 2-10), relative V-Q difference (scale 0.33-1), scaled V-Q difference (scale 0.125-0.32). Bottom row: scaled V-Q difference (scale 0.125-0.32), maximum scaled ventilation and perfusion SPECT. The CT in the latter two images is darker due fusion of the color scales.*

## Discussion

The use of the V-Q difference holds several theoretical and practical advantages over the V/Q ratio for assessing VQ relationships, both in the healthy and the diseased lung. Importantly, an advantage of the scaled difference images is, that the intensity reflects lung function mismatch resulting in another intensity distribution in the images, as illustrated in figure 2-6. Furthermore, both classical mismatches and reversed mismatches can be shown in the same image with the same intensity scale, just with the opposite sign. If a ratio-like intensity scale is required, relative difference images are preferred over ratio images, since classical mismatches and reversed mismatches can be displayed simultaneously, while the scale contrary the classical and logarithmic ratio images is on a closed domain.

Logarithmic ratio and relative difference images look very similar, when both displayed with an intensity scale corresponding to $R$-values up to five, because half the logarithmic ratio and the relative difference defined by equation (3) are equal to second order for deviations from $R$=1 and do not differ more than 10% up to a ratio of three. (Logarithmic) ratio and relative difference images need to be masked, since their values can become very large, for small values of perfusion and the total signal as can be seen in equation (1) and (3). An important advantage of the scaled difference images is, that masking is not necessary for displaying purposes.

When determining a value for a Volume of Interest, the mean of the scaled difference is identical to the scaled difference calculated by the mean perfusion and ventilation values. This is not the case for the ratio and relative difference, where the mean of the ventilation and perfusion need to be calculated first.

Based on our limited experience, a threshold value of 0.125 for maximum scaled difference images seems reasonable, but it might turn out that a slightly different value is more appropriate. This is also the case for the arbitrary threshold of two for the V/Q ratio. This should be investigated in a clinical study, where the difference methods are applied. It can be considered to derive lung function metrics from maximum scaled V-Q difference images. One could for instance consider the sum of the scaled difference with a threshold optionally scaled by the total lung volume measured as the number of voxels.

Ratio images are not so sensitive to attenuation, since both numerator and denominator are influenced by it. Therefore, it is of lesser importance that ratio images are corrected for attenuation. For difference images attenuation correction is necessary. Low dose CT is often performed in clinical practice [14], so the difference images can be computed by using CT-based attenuation corrected ventilation and perfusion images.

As long as images are aligned the difference (and ratio) images outlined in this paper can also be calculated for a sequential VQ scan [15] with [$^{99m}$Tc]Tc-MAA and a $^{99m}$Tc-labelled aerosol of carbon micro-particles (Technegas$^{TM}$) or DTPA [16] as is the case for the ratio images [1]. However, aerosols might cause some extra mismatches in the images of patients due to the hot spots in the large airways [6].

*Conclusions*

A novel method for determining functional difference images in ventilation perfusion SPECT was described. In the future, the clinical value of the maximum scaled functional V-Q difference images needs to be investigated in a larger clinical population. We expect that the scaled difference images will improve the quality and facility of VQ scintigraphy-based diagnostics significantly, as they simultaneously can display mismatches and reversed mismatches, and notably provide a better anatomical delineation of such defects.

## Declarations

Ethics approval and consent to participate

The images are part of a research study approved be the Regional Committee on Health Research Ethics of the Capital Region (H-20028792) and performed in accordance with the Helsinki declaration II. Oral and written informed consent was obtained from all subjects involved in the study.

Consent for publication

Not applicable.

Availability of data and material

The datasets used and/or analyzed during the current study are available from the corresponding author on reasonable request in anonymized form.

Competing interests

The authors declare that they have no competing interests.

Funding

Not applicable.

Authors' contributions

RdN developed the relative and scaled difference method. JM & RMGB were responsible for the scanning of the patients and analyzed the images. SLH discussed alternative measures with RdN at an early stage and contributed with optimization of the CT-scans. All have discussed the newly developed measures intensively. All authors have read, approved and have contributed to the final manuscript.